\documentclass[11pt,a4paper]{article}
\usepackage{jinstpub} 

\usepackage{booktabs}
\usepackage{siunitx}
\usepackage{longtable}
\usepackage{xcolor}
\usepackage{listings}

\lstdefinestyle{spec}{
  basicstyle=\ttfamily\small,
  columns=fullflexible,
  frame=single,
  rulecolor=\color{black!20},
  backgroundcolor=\color{black!2},
  breaklines=true,
  showstringspaces=false
}

\title{T-DAQ-P: a portable tablet-form multi-stream data acquisition and contextual telemetry platform based on COTS modules and a custom integration layer}

\author[a,1]{D.~Tagnani\note{Corresponding author.}}
\author[b]{M.~Andreotti}

\affiliation[a]{Istituto Nazionale di Fisica Nucleare, Sezione di Roma Tre, Rome, Italy}
\affiliation[b]{Istituto Nazionale di Fisica Nucleare, Sezione di Ferrara, Ferrara, Italy}

\emailAdd{diego.tagnani@roma3.infn.it}

\abstract{
We present T-DAQ-P, a compact and portable data acquisition and telemetry platform designed to support detector deployments in laboratory and field conditions by integrating event streaming, slow-control telemetry, and operator-oriented commissioning tools in a single unit.
The system combines a Raspberry Pi~5 host for multi-stream ingestion, visualization, and storage with an Arduino UNO R4 WiFi microcontroller dedicated to sensor acquisition.
The electronics integrate commercial off-the-shelf (COTS) modules through an engineered interface layer including protected power distribution, logic-level adaptation, and a DB-37 expansion connector exposing I\textsuperscript{2}C/SPI/ADC lines for external instrumentation.
On the telemetry side, the microcontroller firmware performs automatic I\textsuperscript{2}C discovery, conditional sensor handling, and outputs framed ``NMEA-like'' messages with XOR checksum and explicit block delimiters, enabling robust parsing and synchronization on the host.
A finite-state control (STOP/RUN/SIM/INIT) and watchdog-style recovery actions implement resilience under partial hardware availability and field failures.
The host software adopts a multi-process architecture with independent readers for event, telemetry, and actuator streams, message queues for isolation, and coherent time-stamped logging across streams.
A simulation mode and integrated command channel allow end-to-end pipeline verification and rapid commissioning without full sensor availability.
The paper details architecture, electronics integration, firmware protocol, host software design, and functional/engineering validation procedures, providing a reproducible blueprint for portable instrumentation readout and contextual telemetry acquisition.
}


\arxivnumber{arXiv:} 

\begin{document}
\maketitle
\flushbottom

\section{Introduction}
Portable detector deployments and field measurement campaigns often require more than event collection: stable operation depends on contextual telemetry (environmental parameters, position/altitude, power health) and on tooling that supports commissioning, diagnostics, and reproducible logging without additional laboratory infrastructure. T-DAQ-P addresses this need by integrating in a single compact unit (i) an event stream interface to a detector readout device, (ii) a dedicated slow-control microcontroller acquiring heterogeneous sensors, (iii) a host computer handling visualization and storage, and (iv) an expansion interface for external instrumentation. This work focuses on architecture, integration, and functional/engineering validation; physics results obtained with hosted detectors are intentionally deferred to separate dissemination.

\section{System overview}
\begin{figure}[t]
  \centering
  \includegraphics[width=0.90\textwidth]{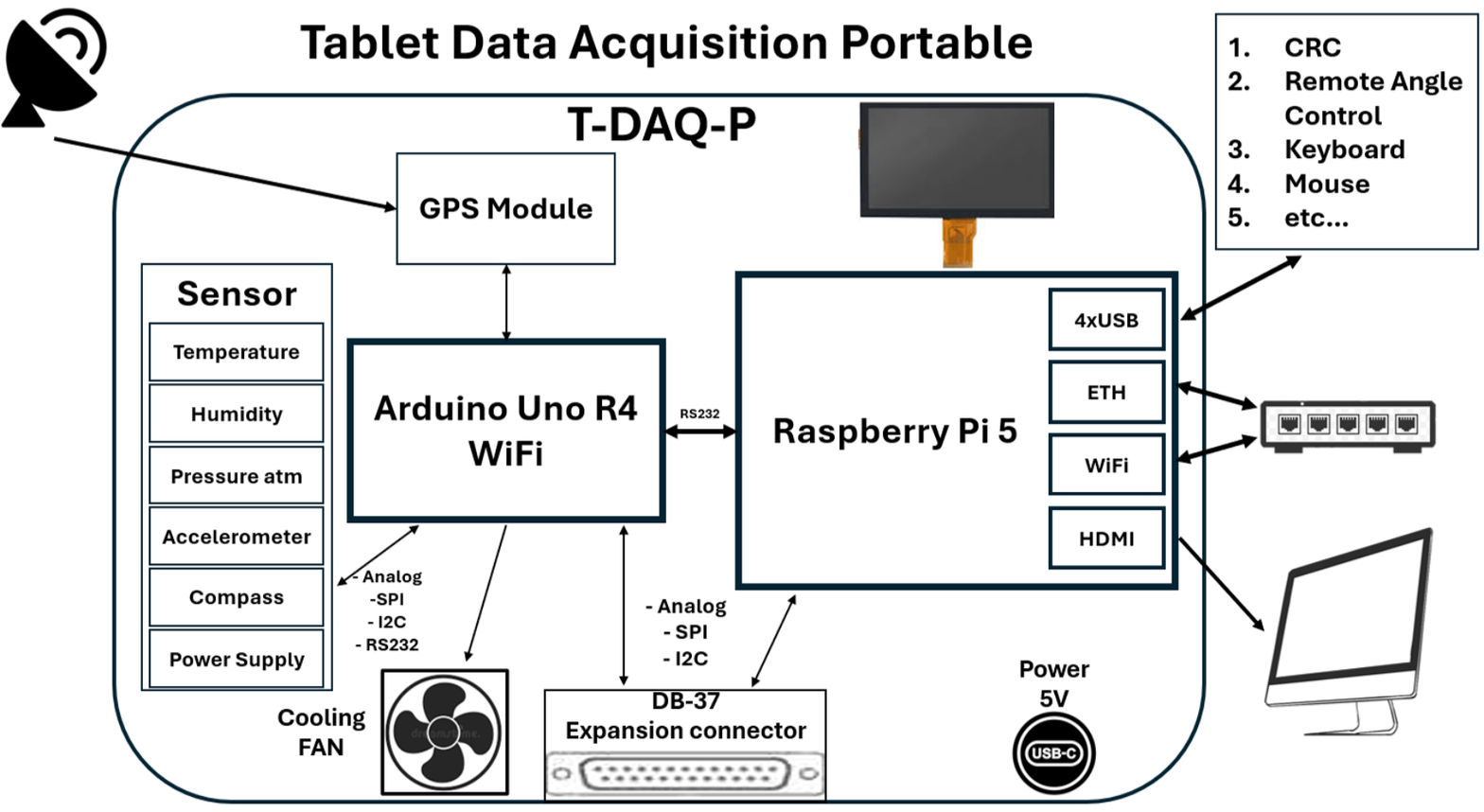}
  \caption{T-DAQ-P system overview (host, integrated local HMI, power entry, thermal management, microcontroller-based slow control, multi-stream serial domains, and expansion interface).}
  \label{fig:system_overview}
\end{figure}

As shown in figure~\ref{fig:system_overview}, T-DAQ-P is organized as a compact two-level acquisition platform that separates (i) host-side multi-stream ingestion, visualization and storage from (ii) microcontroller-side slow-control and contextual telemetry generation. The host layer is implemented on a Raspberry Pi~5, which provides the computing resources and I/O endpoints required for portable deployments, including multiple USB serial connections for external devices, network connectivity, and local storage management~\cite{RPI5_product_brief}. A distinguishing element of the platform, explicitly represented in the block diagram, is the integrated local human--machine interface: a 7$''$ LCD is connected directly to the host and enables stand-alone field operation without requiring an external monitor, while an HDMI output remains available for extended displays when needed~\cite{RPI5_product_brief}. The power entry is likewise treated as a system-level block: the unit is designed around a single external power input via USB-C on the instrument panel, consistent with the Raspberry Pi~5 supply model and supporting a portable ``single-cable'' deployment in the field~\cite{RPI5_product_brief}.

The slow-control layer is implemented on an Arduino UNO R4 WiFi, dedicated to deterministic sensor polling and robust telemetry emission over a UART link to the host~\cite{UNO_R4_docs}. This partitioning reduces coupling between user-facing software and time-critical sensor acquisition, and enables graceful degradation when individual peripherals are unavailable. In addition, the block diagram includes a dedicated thermal-management subsystem (forced-air ventilation through integrated fans and airflow paths), which is relevant for portable enclosures where the host and controller operate continuously under variable ambient conditions; this block is part of the instrument design rather than an external accessory.

The platform is designed to concurrently handle three serial domains: a detector event stream (readout device connected via USB serial), a sensor telemetry stream (microcontroller to host), and an optional actuator/control stream (e.g.\ a remotely controlled support or positioning stage) that may carry both telemetry and command traffic. On the sensing side, the microcontroller integrates heterogeneous devices over a shared I\textsuperscript{2}C bus and dedicated digital/analog inputs: representative sensors include temperature/humidity modules (DHT11/22 and AHT20)~\cite{DHT11_aosong,AHT20_aosong}, barometric pressure/temperature (BMP280)~\cite{BMP280_bosch}, inertial sensing (MPU-6050)~\cite{MPU6050_tdk}, and a 3-axis magnetometer (QMC5883L)~\cite{QMC5883L_qst}. In addition, GNSS timing/position is supported through an external GPS module interfaced to the microcontroller with dedicated RX/TX lines and control pins, allowing the telemetry stream to embed time and georeferencing information in a consistent format~\cite{MIKROE_GPS4Click}.

A key element of figure~\ref{fig:system_overview} is the explicit separation between the \emph{internal} sensing plane and the \emph{external} expansion plane. External instrumentation is supported through a DB-37 connector that exposes a controlled set of expansion resources (power rails, I\textsuperscript{2}C, SPI chip selects and data lines, and ADC channels), enabling additional sensors to be integrated without redesigning the core platform. The DB-37 mapping is treated as part of the hardware interface specification and is summarized in appendix~\ref{app:db37}. At system level, the combination of framed telemetry and process-isolated host ingestion enables reproducible commissioning workflows and coherent multi-stream logging, which are essential in field operation where device availability and connectivity may vary.

\section{Electronics integration and expansion interface}
\label{sec:electronics}
The project is based on a custom integration board that consolidates the COTS elements of T-DAQ-P into a single portable subsystem and formalizes the electrical backbone between the Raspberry Pi~5 host, the slow-control controller, the integrated sensors and the external expansion interface. The schematic makes explicit that the system does not rely on ad-hoc wiring between modules; rather, it defines dedicated power and signal domains, protection elements, mixed-voltage adaptation stages, and a structured breakout of digital buses and analog channels onto a DB-37 connector. In particular, the design introduces distinct power nets for the internal platform and for external loads (e.g.\ dedicated 5~V and 3.3~V rails routed to the expansion connector) and distributes these rails together with multiple ground returns, enabling field extensions while maintaining a clear electrical boundary between the core unit and add-on instrumentation. Protection is implemented at rail level through dedicated fusing elements and local decoupling on the supply branches, an approach intended to bound fault propagation among subsystems and to reduce susceptibility to supply transients when peripherals are connected or when external loads draw unexpected current.

Mixed-voltage integration is treated as a first-class constraint in the board. Since the host platform exposes 3.3~V logic on its GPIO header while the slow-control controller and peripheral ecosystem can involve 5~V domains, the schematic includes explicit logic-level adaptation stages based on MOSFET translators in the BSS138 class, with pull-up resistors on both sides, providing bidirectional level shifting suitable for open-drain buses such as Serial UART and for safe interconnection of mixed-voltage digital nets. This design allows communication between the Raspberry Pi 5 and the Arduino UNO R4 via the hardware serial port on a support board, they work with different digital logic levels.

\begin{figure}[t]
  \centering
  \includegraphics[width=0.90\textwidth]{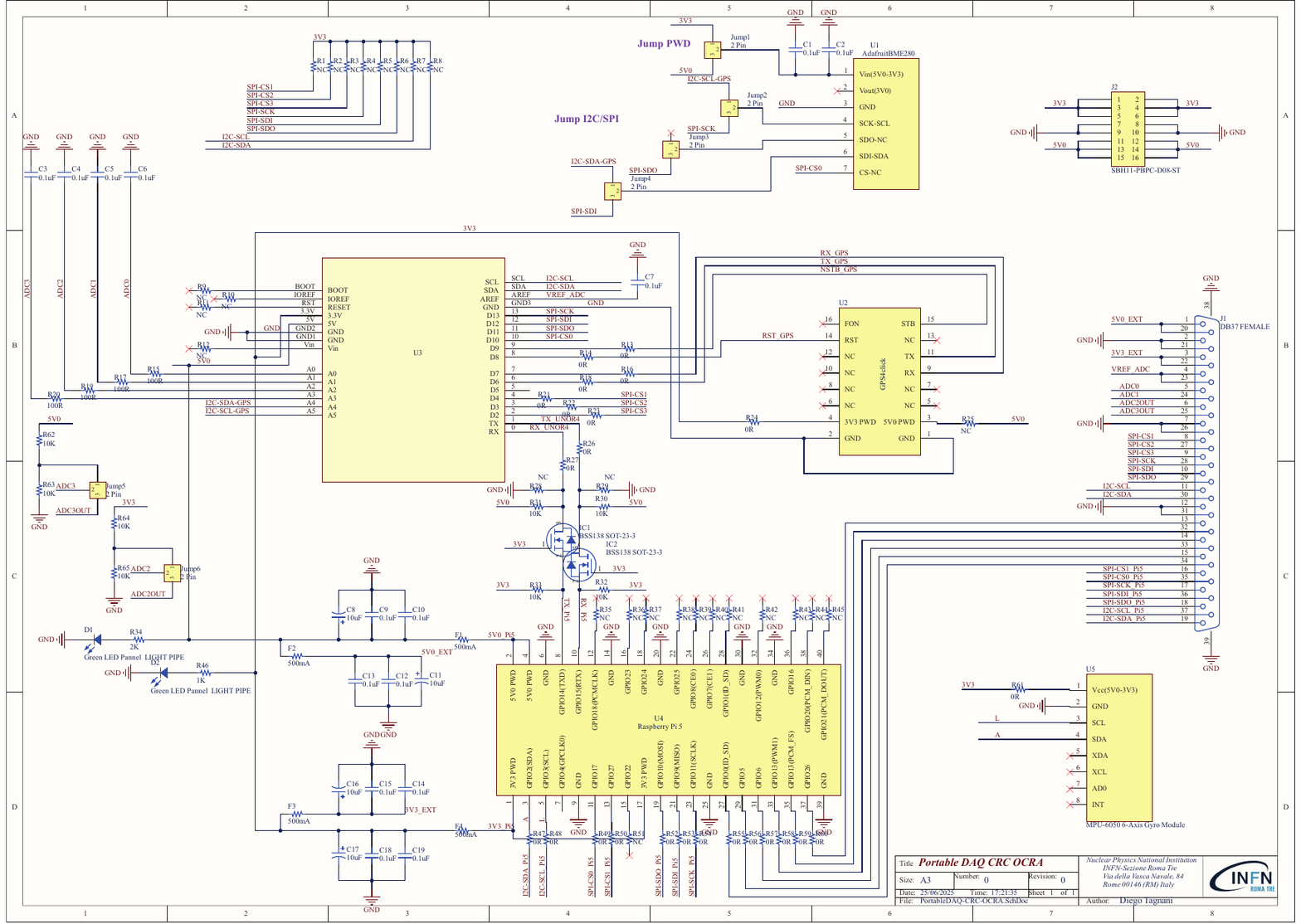}
  \caption{Custom integration board schematic: protected power distribution and mixed-voltage adaptation, shared I\textsuperscript{2}C/SPI sensor backbone, GPS module control, and DB-37 expansion interface exposing power, I\textsuperscript{2}C/SPI and ADC channels.}
  \label{fig:electronics}
\end{figure}

Power entry and protection domains are defined to avoid ambiguity on what is protected by the host platform and what is protected by the custom integration layer. In the reference configuration, the primary power entry is the Raspberry Pi~5 USB-C input operated with a USB Power Delivery supply; the official 27~W PSU provides up to 5.1~V at 5~A and supports PD negotiation~\cite{RPI27W_psu_brief}. When a PD-capable supply is detected, Raspberry Pi~5 can automatically raise the nominal current budget available to the four USB Type-A ports from the default 600~mA to a nominal 1.6~A, enabling higher-power peripherals when the negotiated profile allows it~\cite{RPI27W_psu_brief}. In addition to the USB-C entry, the integration board provides an internal alternative 5~V entry path to the Raspberry Pi~5 5~V header pins through a dedicated fuse (figure~\ref{fig:electronics}). This path is intended for scenarios where the 5~V rail is injected internally rather than through the USB-C PD negotiation path; in this case, the USB-C PD detection mechanism is not exercised and the host follows the default USB peripheral current policy. Downstream fault containment is implemented on the custom board by segmenting loads and applying dedicated fuses to internal peripheral branches and to the DB-37 expansion power rails (figure~\ref{fig:electronics}); this limits fault propagation from external instrumentation (DB-37 side) or internal peripherals to the host power domain, while the Raspberry Pi~5 input stage remains responsible for power-profile negotiation and for host-level current budgeting when powered through USB-C.

At the bus level, figure~\ref{fig:electronics} shows that the integration strategy is based on shared backbones rather than point-to-point wiring: multiple sensors are placed on a common I\textsuperscript{2}C or SPI segment, while the GNSS subsystem is integrated as a managed module with explicit RX/TX signals and dedicated control pins routed on the board. Configuration straps and jumpers provide a controlled way to select routing and operating modes during commissioning and field reconfiguration without altering the core wiring harness. Finally, the DB-37 connector is implemented as an explicit instrumentation interface: it carries power rails, grounds, I\textsuperscript{2}C lines, SPI lines (including multiple chip-selects) and analog channels with a dedicated reference net, providing a clear contract between the core DAQ/slow-control plane and measurement extensions. The pin-level mapping derived from the schematic net labels is reported in appendix~\ref{app:db37}.

\section{Microcontroller firmware: sensing, framed telemetry, and resilience}
\label{sec:firmware}
\begin{figure}[t]
  \centering
  \includegraphics[width=0.92\textwidth]{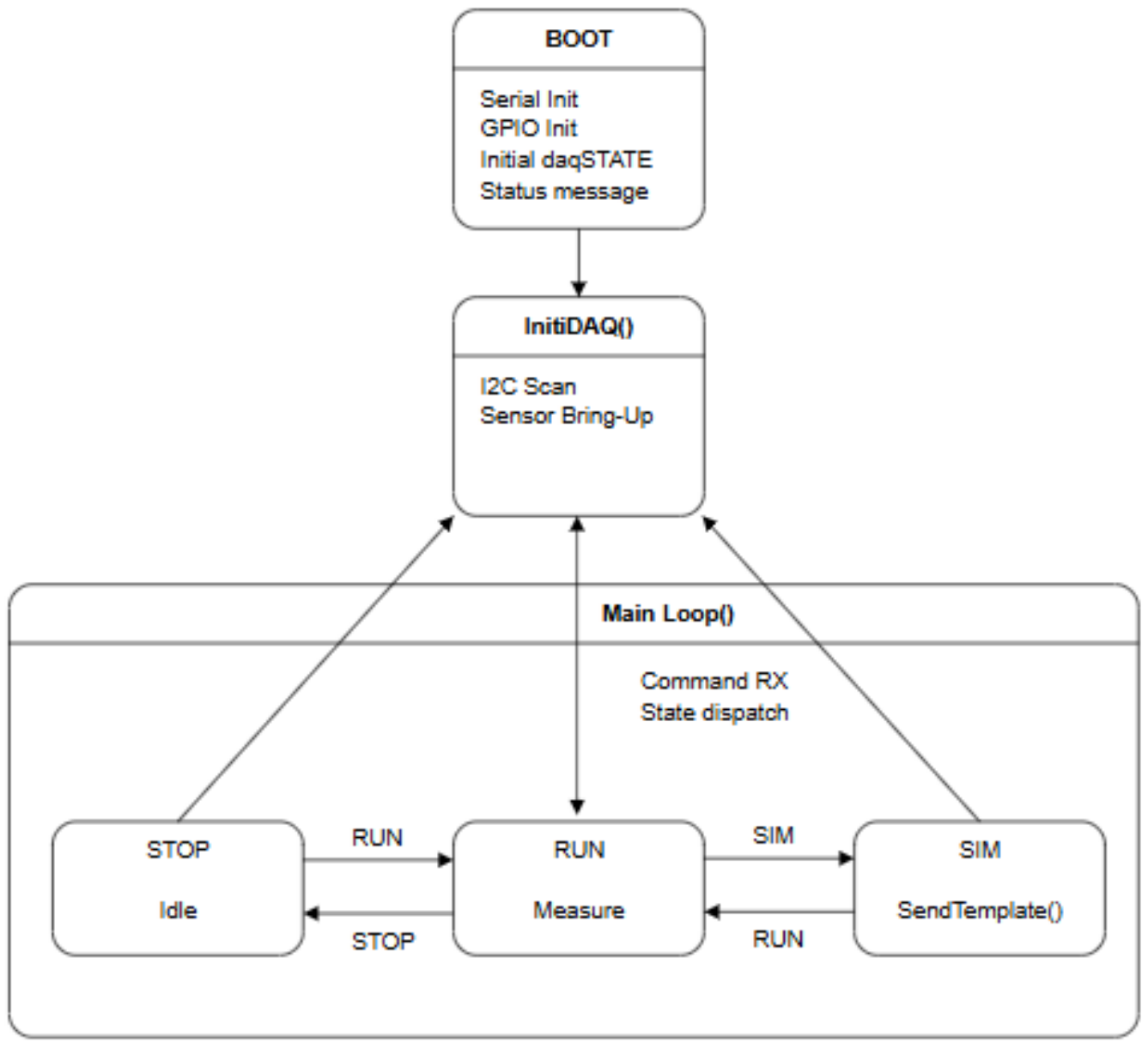}
  \caption{Firmware high-level flow: command reception, state dispatch (STOP/RUN/SIM/INIT), timed acquisition, framed telemetry emission, and resilience hooks.}
  \label{fig:fw_flow}
\end{figure}

The slow-control and contextual telemetry plane of T-DAQ-P is implemented on an Arduino UNO R4 WiFi acting as a dedicated sensor controller and telemetry encoder toward the host, with the explicit goal of keeping the acquisition usable under field constraints such as partial peripheral availability, intermittent GNSS reception, and the need for integrated commissioning without external tooling; the firmware configures a host link on a dedicated hardware UART (Serial1) at 115200~baud and optionally mirrors diagnostic output on the USB serial port when debug is enabled, while GNSS data are received on a dedicated serial interface configured at 9600~baud and coupled to explicit enable/reset control lines so that the GNSS module can be managed by firmware rather than treated as a passive peripheral~\cite{UNO_R4_docs,TinyGPSPlus_repo,TinyGPSPlus_arduino_docs}. The firmware control flow is summarized in figure~\ref{fig:fw_flow}.

The overall structure is organized around a small run-time control state (\texttt{daqState}) that dispatches between STOP, RUN, SIM and INIT behaviors, so that the main loop remains responsive to commands while periodic acquisition is executed on a non-blocking schedule derived from \texttt{millis()}; in the reference configuration provided in the firmware, the telemetry cycle is paced by a configurable period (\texttt{timeloop}) and the GNSS input stream is supervised by a timeout (\texttt{timeout\_gps}) without relying on a fixed \texttt{delay()} in the main loop, preserving command handling and recovery actions in parallel with regular acquisition. A central design choice is automatic discovery and conditional operation of the sensing plane: at initialization the firmware performs an I\textsuperscript{2}C scan over the address space, sets presence flags for supported devices, and conditionally initializes and samples only the sensors detected at run time, allowing the same firmware image to operate across partially populated configurations and enabling graceful degradation when devices are absent or temporarily failing; this includes support for a mixed set of environmental and orientation sensors (temperature/humidity, pressure/temperature, magnetometer/heading, IMU) and also slow-control monitoring of supply rails via analog inputs dedicated to nominal 5~V and 3.3~V measurements, which are converted to engineering units through fixed scaling factors and emitted as part of the telemetry so that the host can track power health alongside environmental context. Telemetry is emitted as a framed, line-oriented stream designed to be both human-inspectable and machine-robust: each acquisition cycle starts with an explicit block marker (\texttt{MEASURE\_START} for real measurements or \texttt{MEASURE\_START\_SIM} for simulated output), then transmits a set of ``NMEA-like'' ASCII sentences with dedicated identifiers and an XOR checksum encoded in hexadecimal appended after a '*' delimiter, computed over the characters between '\$' and '*' to enable sentence-level integrity checking, and finally closes the block with a state/status line (including the run state and initialization counters) and a terminator line \texttt{END\_OF\_MESSAGE}, allowing the host to buffer until completion and process multi-sentence cycles atomically. The command interface is intentionally minimal and commissioning-oriented: newline-terminated textual commands received from the host (STOP, RUN, SIM and INIT) trigger state transitions and are acknowledged through explicit status messages, so that the host application can expose an operator-facing commissioning loop (RUN/SIM/RE-INIT) without a separate terminal; SIM mode is particularly relevant for instrument commissioning because it generates a complete, syntactically valid telemetry block with coherent sentences and checksums, enabling end-to-end validation of serial ingestion, parsing, GUI update and logging independently of sensor availability. Resilience mechanisms are implemented for the two dominant field failure modes addressed by the firmware: GNSS reception loss and sensor-plane degradation; for GNSS, the firmware monitors the absence of meaningful incoming characters and triggers a reset sequence through the GNSS reset pin when a timeout condition is met, using a timed deassert/assert procedure that avoids permanently blocking the acquisition loop, and it reports reset start/completion through the log channel; for the sensing plane, the firmware maintains a per-cycle count of successfully sampled sensor channels and compares it to a configured expectation, triggering a controlled re-initialization sequence when the count falls below threshold, with the number of re-initialization attempts bounded by a maximum, thereby providing recovery from transient I\textsuperscript{2}C faults or disconnects without requiring a full system power cycle. Since this subsystem is intended for slow-control and contextual monitoring rather than high-rate waveform acquisition, the telemetry update rate is kept deliberately low and is bounded by a configurable firmware period. The design target is up to the order of $10^{2}$~Hz (hundreds of Hz at most), with typical deployments operated at lower rates depending on the sensor mix, serial overhead, and logging requirements.

\section{Host software}
The host-side software of T-DAQ-P is implemented in Python~3 and is designed to integrate acquisition, monitoring, control, and logging of multiple independent serial subsystems while remaining responsive in a portable ``acquisition tablet'' configuration. The implementation adopts a process-isolated architecture based on \texttt{multiprocessing}, where each hardware subsystem is handled by a dedicated acquisition process and the graphical user interface acts as a supervisory layer consuming decoded records through inter-process queues rather than performing blocking serial I/O directly; this choice improves fault containment when a single endpoint becomes temporarily unavailable and preserves operator feedback during commissioning and field operation~\cite{Python_multiprocessing}. Serial transport is implemented with \texttt{pySerial} and, in the reference configuration, three concurrent streams are ingested: the detector event stream (CRC), the microcontroller telemetry stream (environmental and status information framed by \texttt{END\_OF\_MESSAGE}), and the auxiliary Culla stream, each with its own buffering strategy to reconstruct complete logical messages before updating the GUI or writing to disk~\cite{pySerial_docs}. Bidirectional control is supported through command queues for the microcontroller and for the Culla subsystem, enabling integrated commissioning actions (e.g.\ \texttt{run}/\texttt{sim}/\texttt{init} toward the sensor controller) and setpoint transmission toward the auxiliary platform without external terminal tools. The GUI is implemented with Tkinter and organizes operational monitoring and debugging views in a tab-based layout; in addition, the muon counting rate history is rendered with Matplotlib embedded into the Tkinter canvas, with bounded buffers used for both event storage and plotting to prevent uncontrolled memory growth during long acquisitions~\cite{Python_tkinter,Matplotlib_docs}. In the current implementation, queue polling is performed on a 200~ms cadence and plot refresh is performed every 2~s, while the rate computation uses an operator-selectable gate window (30--240~s) and derives the displayed rate from the number of events falling inside the most recent gate interval. Data logging is controlled by shared software flags and is executed within acquisition processes to minimize UI-induced latency; daily files are produced per subsystem (CRC events, derived CRC rate, microcontroller telemetry, Culla telemetry) with a consistent naming scheme and an optional run suffix, and each entry is time-tagged to support a posteriori merging of independently written data products. A key design choice for synchronization is that the Raspberry Pi~5 assigns a host-side timestamp to each received record upon parsing, and this host timestamp is used as the common timebase across all generated files; when available, the Raspberry Pi~5 real-time clock can be battery-backed to preserve time across power cycles~\cite{RPI5_rtc_battery}, and GNSS-provided UTC information embedded in the telemetry stream can be used for coherence checks between host time and GNSS time and as an independent absolute reference when GNSS lock is available~\cite{MIKROE_GPS4Click}.

\section{Dispatcher: Serial stream virtualization for concurrent consumers}
\begin{figure}[t]
  \centering
  \includegraphics[width=0.85\textwidth]{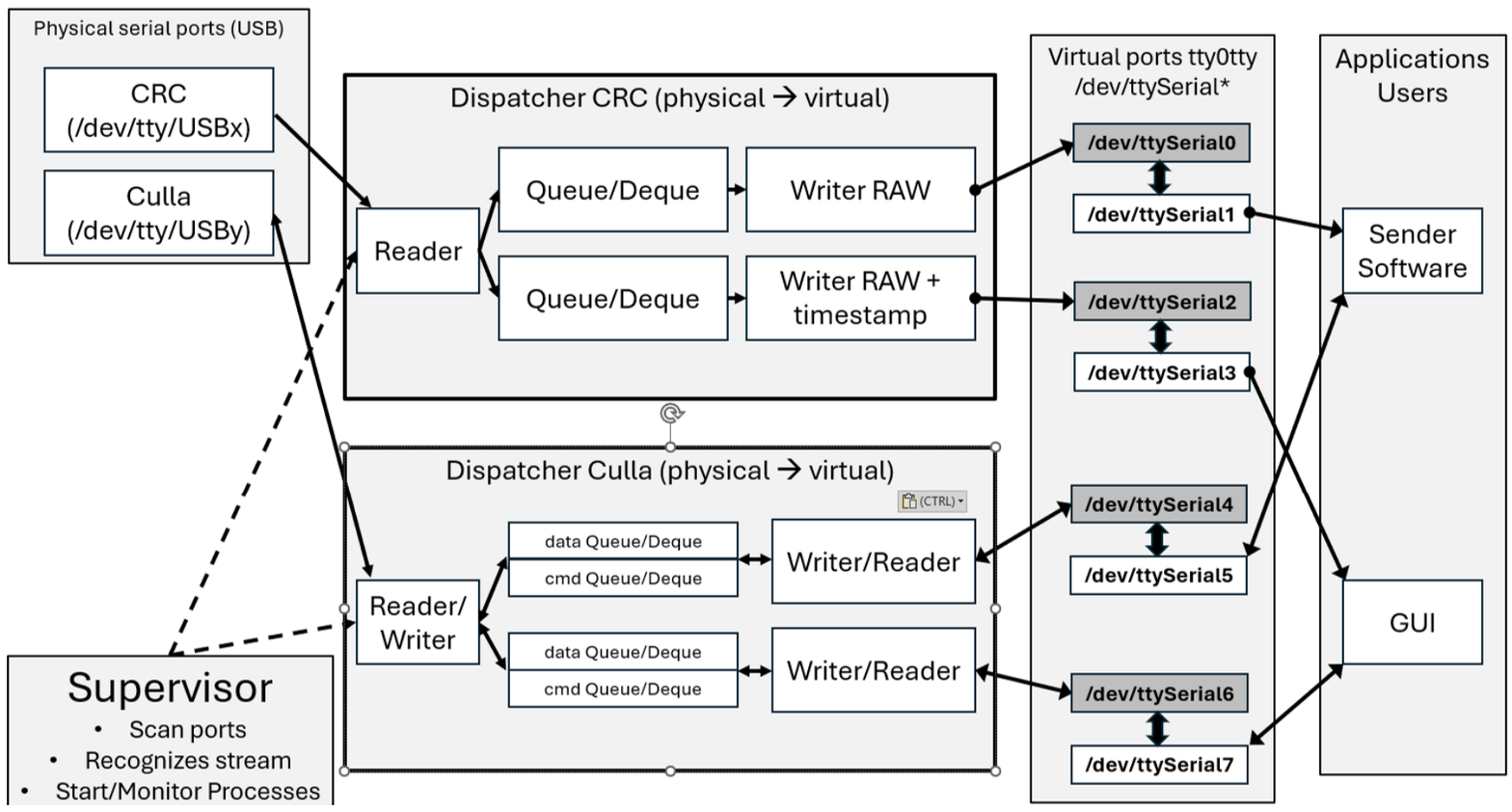}
  \caption{Conceptual serial stream virtualization: physical device stream replicated/bridged to virtual ports to support concurrent consumers.}
  \label{fig:dispatcher}
\end{figure}

In portable deployments it is often necessary to provide the same physical serial stream to more than one software consumer, for instance to allow simultaneous monitoring and forwarding or independent logging and visualization pipelines; however, a physical serial endpoint can typically be opened by a single client at a time. To support this workflow, T-DAQ-P adopts a serial stream virtualization layer implemented as a system service (Serial Dispatcher) that redirects selected physical streams to multiple virtual serial endpoints, enabling concurrent consumers without contention. The virtualization is realized by creating paired virtual serial ports via the Linux kernel null-modem emulator \texttt{tty0tty}, and by running a Python supervisor that detects connected serial devices, classifies incoming streams by inspecting their content, and launches the appropriate dispatcher processes to replicate or bridge the traffic~\cite{tty0tty_repo,Python_multiprocessing}. In the CRC use case, the dispatcher replicates the incoming detector stream onto two virtual ports so that independent applications can consume the same data in parallel, while preserving the ability to augment one copy with host-side metadata (e.g.\ timestamps) when required by a downstream pipeline; in the Culla use case, the dispatcher provides a bidirectional bridge between the physical device and the virtual endpoints so that both telemetry and command traffic can be handled transparently. Robustness is achieved by supervising the dispatcher processes and re-establishing routing after disconnections or timeouts, so that the service can run continuously without manual intervention during field campaigns. A conceptual view of the virtualization scheme is shown in figure~\ref{fig:dispatcher}.

\section{Commissioning (ongoing) -- Preliminary Results}
\label{sec:commissioning_prelim}
Commissioning activities are currently in progress; therefore, the results reported here must be regarded as preliminary. To enable a uniform quantitative comparison across heterogeneous monitored quantities, we report a relative \emph{precision} metric defined as $P = 100 \cdot \sigma_{\mathrm{short}}/\mu$ (in \%), where $\mu$ is the mean value over the considered acquisition and $\sigma_{\mathrm{short}}$ is a short-term repeatability estimator consistent with the rolling standard deviation displayed as error bars in the plots. We intentionally do not quote a long-horizon stability figure at this stage, because the commissioning data were acquired under non-stationary environmental conditions (true temperature/humidity/pressure changes), which would dominate any excursion-based metric and make it instrument-non-specific; long-term stability will be extracted in a future revision using dedicated stationary tests.

Using the above definition, the co-located temperature channels show a relative agreement at the level of $P_T \simeq 0.7$--$1.1\,\%$, while relative humidity exhibits $P_{\mathrm{RH}} \simeq 0.9$--$1.6\,\%$. The pressure channel provides the best short-term repeatability, with $P_{p} \simeq (5$--$10)\times10^{-3}\,\%$ under the tested conditions. For altitude, the median-filtered GPS estimate reduces the scatter with respect to the raw solution and remains consistent with the barometric proxy (inverse-pressure trend after linear rescaling), yielding a figure-extracted repeatability of $P_{h} \simeq 1$--$2\,\%$. Final values will be consolidated after analysis of the full raw dataset and after dedicated controlled-condition runs.

During field operation the unit is either fully active (measurement mode) or powered off. Using a 12~V, 60~Ah battery pack, an autonomy of up to approximately 24~h was observed, corresponding to a nominal available energy of about $E \simeq 12~\mathrm{V}\times60~\mathrm{Ah}\approx720~\mathrm{Wh}$ and an average battery-side current of $I_{\mathrm{avg,batt}}\approx60~\mathrm{Ah}/24~\mathrm{h}\approx2.5~\mathrm{A}$ (system-level figure). When the unit is powered through the Raspberry Pi~5 USB-C input, the internal 5~V domain is bounded by the negotiated USB-C PD profile (up to 5.1~V at 5~A for the official PSU); therefore the effective battery-side power draw depends on the power-conversion chain and on the complete load set (host, display, peripherals, and expansion loads) rather than on the host alone~\cite{RPI27W_psu_brief}.

\begin{figure}[t]
  \centering
  \includegraphics[width=\linewidth]{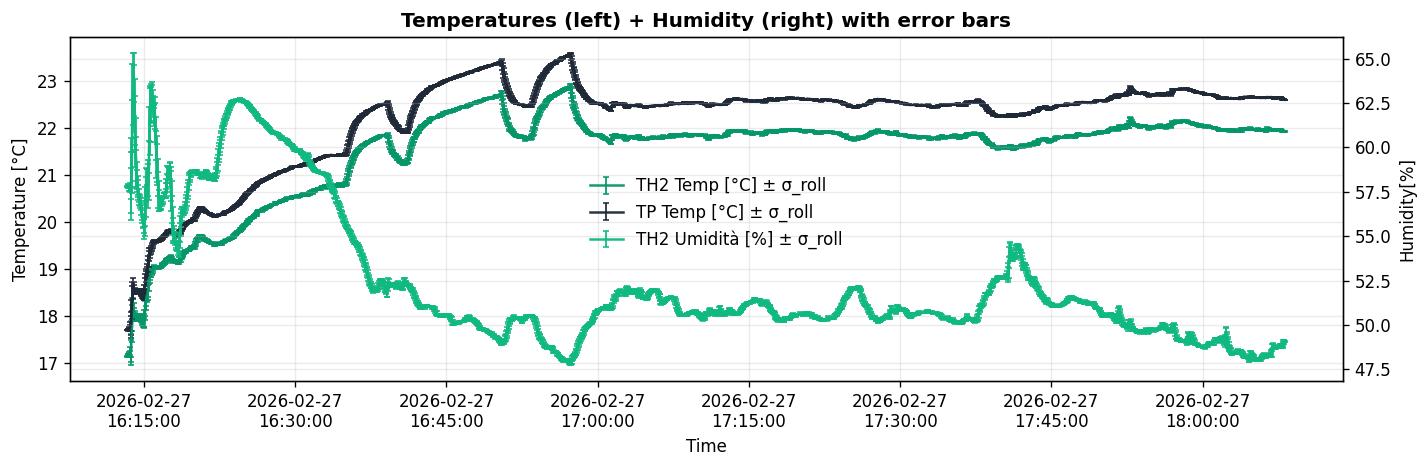}
  \caption{Environmental monitoring channels: temperature measurements (left axis) and relative humidity (right axis) with rolling-standard-deviation error bars; the lower panel shows the atmospheric pressure time series.}
  \label{fig:env_temp_rh_pressure}
\end{figure}

\begin{figure}[t]
  \centering
  \includegraphics[width=\linewidth]{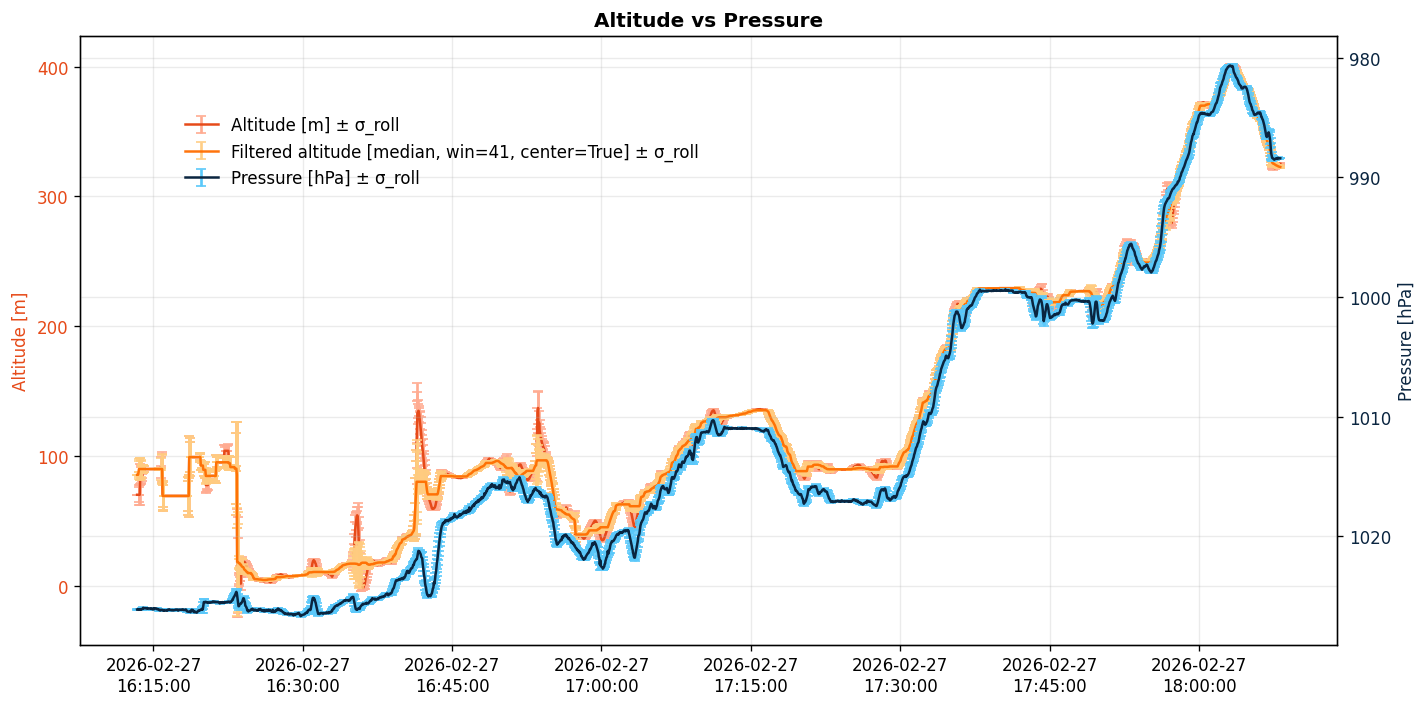}
  \caption{Altitude reconstruction compared with atmospheric pressure. The GPS-derived altitude is shown both as raw and median-filtered solution, while pressure is plotted on an inverted scale to highlight the expected anti-correlation. Error bars represent the rolling dispersion.}
  \label{fig:altitude_vs_pressure}
\end{figure}

\begin{figure}[t]
  \centering
  \includegraphics[width=\linewidth]{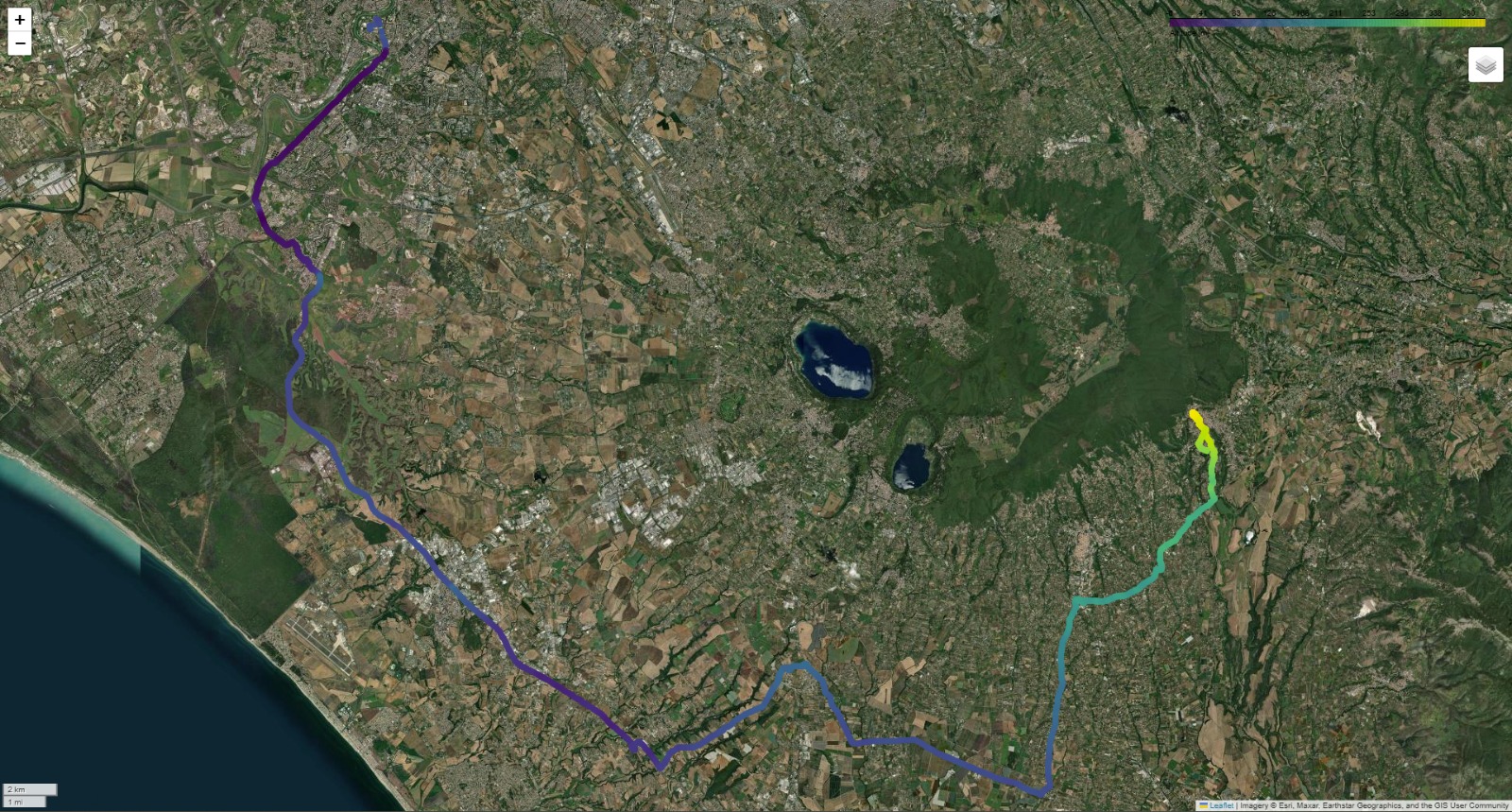}
  \caption{Geographical overview of the measurement route used during commissioning, shown for qualitative context of the GNSS-based altitude and track reconstruction.}
  \label{fig:route_overview}
\end{figure}

\section{Discussion}
T-DAQ-P is intentionally designed around a hybrid philosophy: leveraging widely available commercial off-the-shelf (COTS) computing and sensing modules, while enforcing system-level coherence through a dedicated custom integration layer (figure~\ref{fig:electronics}) and a well-defined software/telemetry contract. The key point is not merely that COTS parts are convenient, but that they become \emph{instrumentation-grade} when their electrical domains, protection strategy, bus routing, and data interfaces are formalized in a repeatable way. In this sense, the custom board is not an ``extra PCB'': it acts as a stable backplane that defines power distribution boundaries, mixed-voltage interoperability, expansion resources (DB-37), and a controlled attachment point for external instrumentation, while keeping the host and slow-control boards replaceable as modules.

Host modularity is constrained by the custom integration layer being electrically and mechanically coupled to the Raspberry Pi 40-pin GPIO header; host substitutions therefore require a pin-compatible Raspberry Pi-class form factor or a revised carrier/integration PCB.

This approach has practical consequences for the lifecycle of a portable DAQ. COTS single-board computers and microcontroller platforms are maintained by their vendors with continuous documentation updates, product briefs, and ecosystem evolution; for example, Raspberry Pi provides consolidated product briefs and Arduino provides board-specific reference documentation for the UNO R4 WiFi platform~\cite{RPI5_product_brief,UNO_R4_docs}. In a system where the custom integration layer defines the electrical contract (rails, level shifting, connector pinout) and the firmware/host interface defines the data contract (framing, checksums, state control), future upgrades can be approached primarily as \emph{module substitutions}: adopting a more capable host board or a revised controller board from the same ecosystem may require limited changes in mounting, cabling, and software configuration, while preserving the validated custom integration and the established acquisition workflow. This reduces the need to repeatedly redesign and reproduce the custom hardware each time a vendor releases an updated COTS platform, provided that the relevant mechanical and electrical interfaces are preserved within the chosen ecosystem.

From an instrumentation standpoint, the most valuable aspect of this ``stable custom layer + evolving COTS'' strategy is risk containment. The custom board concentrates the parts that are expensive to re-qualify in the field: protection and power segmentation, mixed-voltage adaptation, and the external expansion contract. Conversely, the parts that benefit most from vendor-driven improvements (compute performance, I/O bandwidth, maintenance of operating system support, and availability of development tools) remain modular. The same reasoning applies to the software side: once the framed telemetry protocol and the multi-stream ingestion model are in place, commissioning and logging workflows remain invariant even if the underlying host platform is updated, because the host application expects a stable framed stream and exposes the same operational hooks (run/sim/init) to the slow-control firmware.

Finally, the described architecture is deliberately aligned with realistic field constraints: partial peripheral availability, intermittent GNSS reception, and the need for rapid diagnosis without external tools. Simulation mode, explicit state reporting, per-stream freshness indicators, and coherent multi-stream logging are not ``user interface features'' but engineering mechanisms that shorten commissioning cycles and preserve traceability under non-ideal conditions. Overall, T-DAQ-P demonstrates a portable DAQ design pattern where COTS modules provide a sustainable upgrade path, while a custom-engineered integration layer and a strict data/interface contract preserve measurement context, robustness, and reproducibility over the instrument lifetime.

\section{Conclusions}
We presented T-DAQ-P, a miniaturized portable DAQ and telemetry platform integrating a host computer and a dedicated slow-control microcontroller, engineered electronics for COTS module integration, a framed NMEA-like telemetry protocol with checksum, resilience mechanisms for field operation, and a multi-process host architecture supporting monitoring, commissioning, and coherent logging of heterogeneous streams. The system provides a reproducible blueprint for portable instrumentation readout contexts where contextual telemetry and robust operation are as important as event collection. The complete instrument has been realized as an integrated portable unit, including a dedicated mechanical enclosure that consolidates the computing, slow-control, power entry, thermal management, and user-interface elements into a field-deployable ``acquisition tablet''. Commissioning activities are currently ongoing and have already demonstrated robust end-to-end operation of the acquisition, telemetry framing, control, and logging chain in realistic deployment conditions.

\acknowledgments
The authors acknowledge the colleagues of the INFN OCRA project for their continued collaboration in the scientific measurement activities and in the outreach programme with schools based on the Cosmic Ray Cube (CRC). We also thank Digitalcomoedia S.r.l. for the development and maintenance of the Linux/Windows sender/receiver software stack and for its adaptation to the Raspberry Pi platform used in T-DAQ-P. This work was supported by INFN Third Mission initiatives and by INFN Sezione di Roma Tre.

\appendix
\section{DB-37 expansion connector pin mapping}
\label{app:db37}
Table~\ref{tab:db37} reports the DB-37 pin-level mapping derived from the schematics.

\begin{longtable}{@{}r l@{}}
\caption{DB-37 expansion connector mapping (pin $\rightarrow$ signal).}\label{tab:db37}\\
\toprule
Pin & Signal \\ \midrule
\endfirsthead
\toprule
Pin & Signal \\ \midrule
\endhead
\midrule
\multicolumn{2}{r}{\emph{Continued on next page}}\\
\endfoot
\bottomrule
\endlastfoot

 1  & \texttt{5V0\_EXT} \\
 2  & \texttt{GND} \\
 3  & \texttt{3V3\_EXT} \\
 4  & \texttt{VREF\_ADC} \\
 5  & \texttt{ADC0} \\
 6  & \texttt{ADC2OUT} \\
 7  & \texttt{GND} \\
 8  & \texttt{SPI-CS1} \\
 9  & \texttt{SPI-CS3} \\
10  & \texttt{SPI-SDI} \\
11  & \texttt{I2C-SCL} \\
12  & \texttt{GND} \\
13  & \texttt{Pi\_GPIO26} \\
14  & \texttt{Pi\_GPIO13} (PWM1) \\
15  & \texttt{Pi\_GPIO5} \\
16  & \texttt{Pi\_SPI-CS1} \\
17  & \texttt{Pi\_SPI-SCK} \\
18  & \texttt{Pi\_SPI-SDO} \\
19  & \texttt{Pi\_I2C-SDA} \\
20  & \texttt{5V0\_EXT} \\
21  & \texttt{GND} \\
22  & \texttt{3V3\_EXT} \\
23  & \texttt{VREF\_ADC} \\
24  & \texttt{ADC1} \\
25  & \texttt{ADC3OUT} \\
26  & \texttt{GND} \\
27  & \texttt{SPI-CS2} \\
28  & \texttt{SPI-SCK} \\
29  & \texttt{SPI-SDO} \\
30  & \texttt{I2C-SDA} \\
31  & \texttt{GND} \\
32  & \texttt{Pi\_GPIO13} \\
33  & \texttt{Pi\_GPIO6} \\
34  & \texttt{Pi\_GPIO0} \\
35  & \texttt{Pi\_SPI-CS0} \\
36  & \texttt{Pi\_SPI-SDI} \\
37  & \texttt{Pi\_I2C-SCL} \\
\end{longtable}

\bibliographystyle{JHEP}
\bibliography{references}

\end{document}